\documentclass[%
 aip,
 amsmath,amssymb,
 reprint,%
]{revtex4-1}

\usepackage{graphicx}
\usepackage{subfigure}
\usepackage{dcolumn}
\usepackage{bm}
\usepackage{xcolor}

\usepackage[utf8]{inputenc}
\usepackage[T1]{fontenc}
\usepackage{mathptmx}
\usepackage{etoolbox}

\newcommand{\citeg}[1]{\citep[e.g.][]{#1}}

\makeatletter
\def\@email#1#2{%
 \endgroup
 \patchcmd{\titleblock@produce}
  {\frontmatter@RRAPformat}
  {\frontmatter@RRAPformat{\produce@RRAP{*#1\href{mailto:#2}{#2}}}\frontmatter@RRAPformat}
  {}{}
}%
\makeatother
\begin{document}

\preprint{AIP/123-QED}

\title{The payload of the Lunar Gravitational-wave Antenna}

\author{J.V. van Heijningen}
\affiliation{Centre for Cosmology, Particle Physics and Phenomenology (CP3), UCLouvain, B-1348 Louvain-la-Neuve, Belgium}
\author{H.J.M. ter Brake}
 \affiliation{Faculty of Science and Technology, University of Twente, 7522 NB Enschede, The Netherlands}
 \author{O. Gerberding}
 \affiliation{Institut für Experimentalphysik, Universität Hamburg, 22761 Hamburg, Germany}
  \author{S. Chalathadka Subrahmanya}
 \affiliation{Institut für Experimentalphysik, Universität Hamburg, 22761 Hamburg, Germany}
  \author{J. Harms}
 \affiliation{Gran Sasso Science Institute (GSSI), I-67100 L’Aquila, Italy}
  \author{X. Bian}
 \affiliation{Institute of Mechanics, Chinese Academy of Sciences, Beijing 100190, China}
 \author{A. Gatti}
\affiliation{ESAT-MICAS, Katholieke Universiteit Leuven, 3001 Leuven, Belgium}
 \author{M. Zeoli}
\affiliation{Centre for Cosmology, Particle Physics and Phenomenology (CP3), UCLouvain, B-1348 Louvain-la-Neuve, Belgium}
\author{A. Bertolini}
 \affiliation{National institute of subatomic physics Nikhef, 1098 XG Amsterdam, The Netherlands}
 \author{C. Collette}
 \affiliation{Precision Mechatronics Laboratory, Université de Liège, B-4000, Liège, Belgium}
\author{A. Perali}%
\affiliation{School of Pharmacy, Physics Unit, University of Camerino, I-62032 Camerino (MC), Italy}
\affiliation{INAF, I-62032 Camerino (MC), Italy}
\author{N. Pinto}%
\affiliation{School of Science and Technology, Physics Division, University of Camerino, I-62032 Camerino (MC), Italy}
\author{M. Sharma}%
\affiliation{School of Science and Technology, Physics Division, University of Camerino, I-62032 Camerino (MC), Italy}
 \author{F. Tavernier}
\affiliation{ESAT-MICAS, Katholieke Universiteit Leuven, 3001 Leuven, Belgium}
\author{J. Rezvani}%
\affiliation{School of Science and Technology, Physics Division, University of Camerino, I-62032 Camerino (MC), Italy}

\date{\today}

\begin{abstract}
The toolbox to study the Universe grew on 14 September 2015 when the LIGO--Virgo collaboration heard a signal from two colliding black holes between 30-250\,Hz. Since then, many more gravitational waves have been detected as detectors increased sensitivity. However, the current detector design sensitivity curves still have a lower cut-off of 10\,Hz. To detect even lower-frequency gravitational-wave signals, the Lunar Gravitational-wave Antenna will use an array of seismic stations in a permanently shadowed crater. It aims to detect the differential between the elastic response of the Moon and the suspended inertial sensor proof mass motion induced by gravitational waves. A cryogenic superconducting inertial sensor is under development that aims for fm/$\surd$Hz sensitivity or better down to 1\,Hz and is planned to be deployed in seismic stations. Here, we describe the current state of research towards the inertial sensor, its applications and additional auxiliary technologies in the payload of the lunar gravitational-wave detection mission.

\end{abstract}

\maketitle

The future of gravitational waves (GWs) is bright. After the first detection of a binary black hole  merger in 2015\,\cite{GW150914} and a binary neutron star merger with electromagnetic counterpart in 2017\,\cite{GW170817}, the LIGO-Virgo-KAGRA collaboration has detected more than 90 signals from black hole and/or neutron star mergers in their first 3 observation runs\,\cite{GWTC3} using the LIGO\,\cite{aLIGO} and Virgo\,\cite{AdVirgo} detectors. KAGRA\,\cite{KAGRA}, the first underground and cryogenic detector, will join in the coming observation run. All measured signals entered the LIGO/Virgo sensitive band at around 30\,Hz. Technical noise from many cross couplings between angular and translational control, is the dominant noise source below 30\,Hz. By improving the low-frequency performance, signals could be longer in-band and we could have access to a population of BBH systems with a total mass greater than 200\,$M_{\odot}$. 

The Lunar GW Antenna (LGWA)\,\cite{LGWA} will detect GWs in the decihertz region (0.1 -- 1\,Hz), giving access to even more massive BBH systems, white dwarf binaries and tidal disruption events such as a star plunging into a black hole. LGWA uses an array of extremely sensitive inertial sensors to probe directly the deformation of the lunar body as a result of the passing GW. In summary, the lunar surface -- and the rigidly attached inertial sensor suspension frame -- displaces according to an elastic response determined by the stiffness of the lunar body; the proof mass of the inertial sensor, however, displaces inertially and so the differential displacement between proof mass and suspension frame holds the GW signal. More details on this detection principle are found in ref.\,\cite{Harms2022}.

First, the mission concept is described in section~\ref{MisConc}, focusing on the heart of the antenna: the seismic station. In order to achieve sufficient sensitivity to strain, we propose using an array of high-performance inertial sensors; section~\ref{Inert} describes the development of such (sub-)fm/$\surd$Hz class inertial sensors. A necessity to reach such sensitivity also down to low frequency is the use of cryogenics which will lower thermal noise and enable the use of high-$Q$ superconducting actuation and possibly sensing; sorption cooling and thermal management is described in section~\ref{SorpTherm}. High mechanical sensitivity and low thermal noise are obtained by extremely soft proof mass suspension. This sets strict requirements on the leveling system, described in section\,\ref{Level}. Finally, we detail the synergy of LGWA inertial sensor development with the next-generation terrestrial GW detector Einstein Telescope (ET) in section~\ref{Syn}.

\section{Mission concept and seismic stations}\label{MisConc}
The Lunar GW Antenna is a proposed kilometer-scale array of four seismic stations deployed on the lunar surface. Each station measures the horizontal surface displacement along two orthogonal directions. The horizontal direction is chosen to be able to build softer proof-mass suspensions, which benefits the instrument sensitivity (see following sections). The LGWA deployment site is one of the permanently shadowed regions inside a crater at the lunar north or south pole. Without direct sunlight, alternatives to solar panels on our stations are investigated. One of the possible power system for LGWA laser-power beaming system using solar panels on the crater edge\,\cite{GlEA2018}.

\begin{figure}[htb]
    \centering
    \includegraphics[width = 0.98\linewidth]{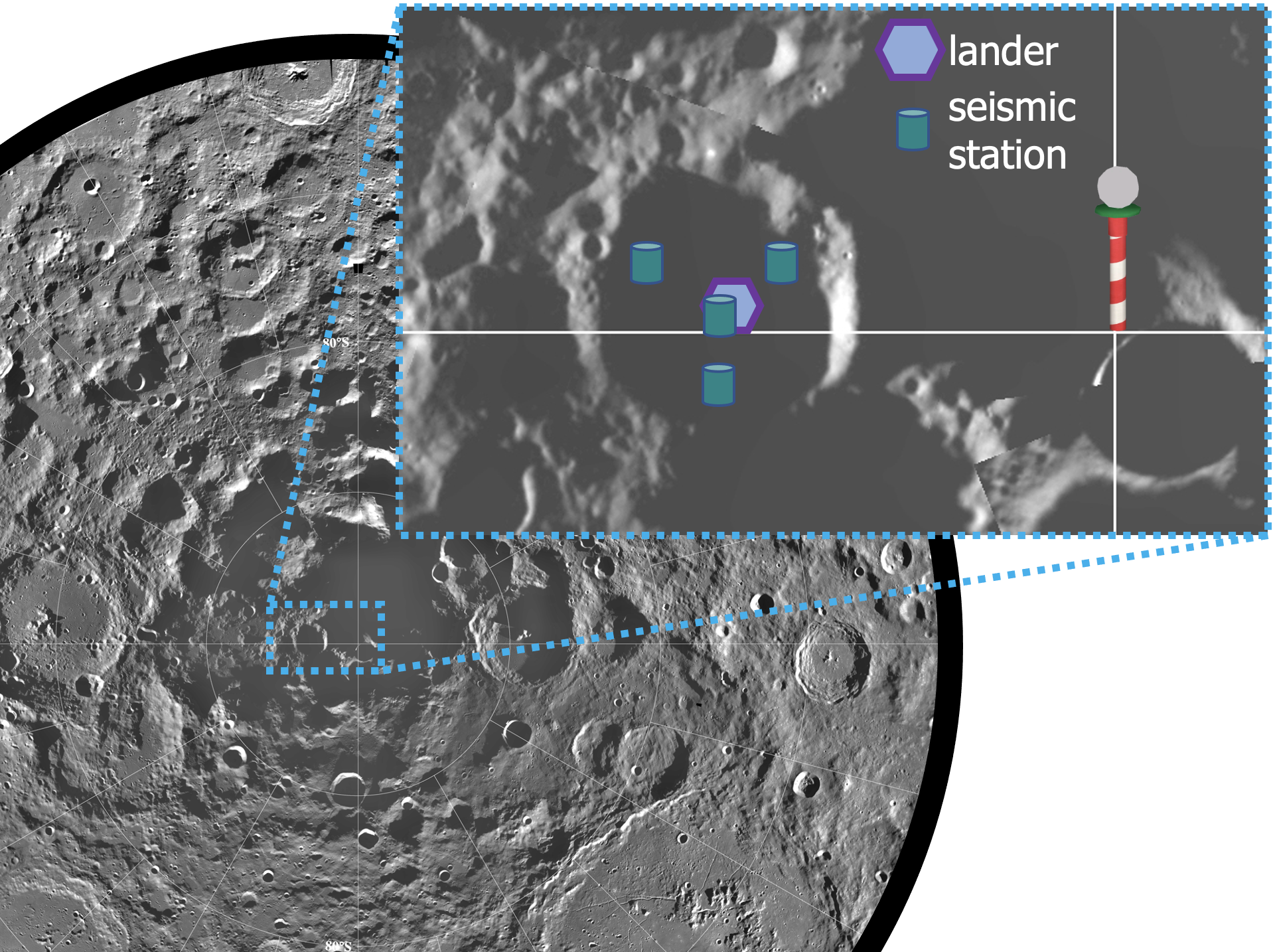}
    \caption{Lunar mosaic of about 1500 Clementine images of the lunar south polar region. The projection is orthographic, centred on the south pole out to 70$^o$ S. The Schr\"{o}dinger Basin (320 km in diameter) is located in the lower right. The inset shows an example crater near the south pole with a star-like deployment configuration of a lander and four seismic stations in a kilometer-scale array of seismic station containing cryogenic inertial sensors. Adapted from ref.\,\cite{Mosaic}.}
    \label{fig:crater}
\end{figure}

While each seismometer has the capability to observe a GW signal, the array is proposed as a tool for the reduction of the seismic background in LGWA data. The models of the seismic background still need to be improved, but the preliminary results indicate that a background limitation of GW measurements with LGWA should be expected above 0.1\,Hz \cite{LGWA,Lognonne2009,Harms2022}. Work is underway to generalize noise-cancellation methods developed for current GW detectors \cite{Badaracco2020} to be applicable to LGWA. The star-like array configuration shown in figure \ref{fig:crater} is proposed with the idea to achieve best noise cancellation in the central sensor. 

Crucial for the success of LGWA is the excellent quality of the Moon as ultra-quiet elastic body responding to the extremely weak spacetime fluctuations. The lunar seismic background from meteoroid impacts is predicted to be several orders of magnitude quieter than the terrestrial seismic background \cite{Lognonne2009}. Other sources of surface displacement must generally be considered. Albeit higher in magnitude when compared to other types of moonquakes, shallow moonquakes are rare and not expected to significantly reduce observation time of lunar GW detectors. Deep moonquakes are more frequent, but the corresponding background noise is expected to lie below the one from meteoroid impacts. Also thermal effects can lead to seismic events. The so-called thermal moonquakes were observed in large numbers with the Apollo seismic stations around sunset and sunrise \cite{DuSu1974}. It is also to be expected that temperature changes lead to ground tilts and deformations of payload and lander causing additional disturbances of seismic measurements \cite{StEA2020}. 

\begin{figure}[h]
  \centering
    \includegraphics[width = 1\linewidth]{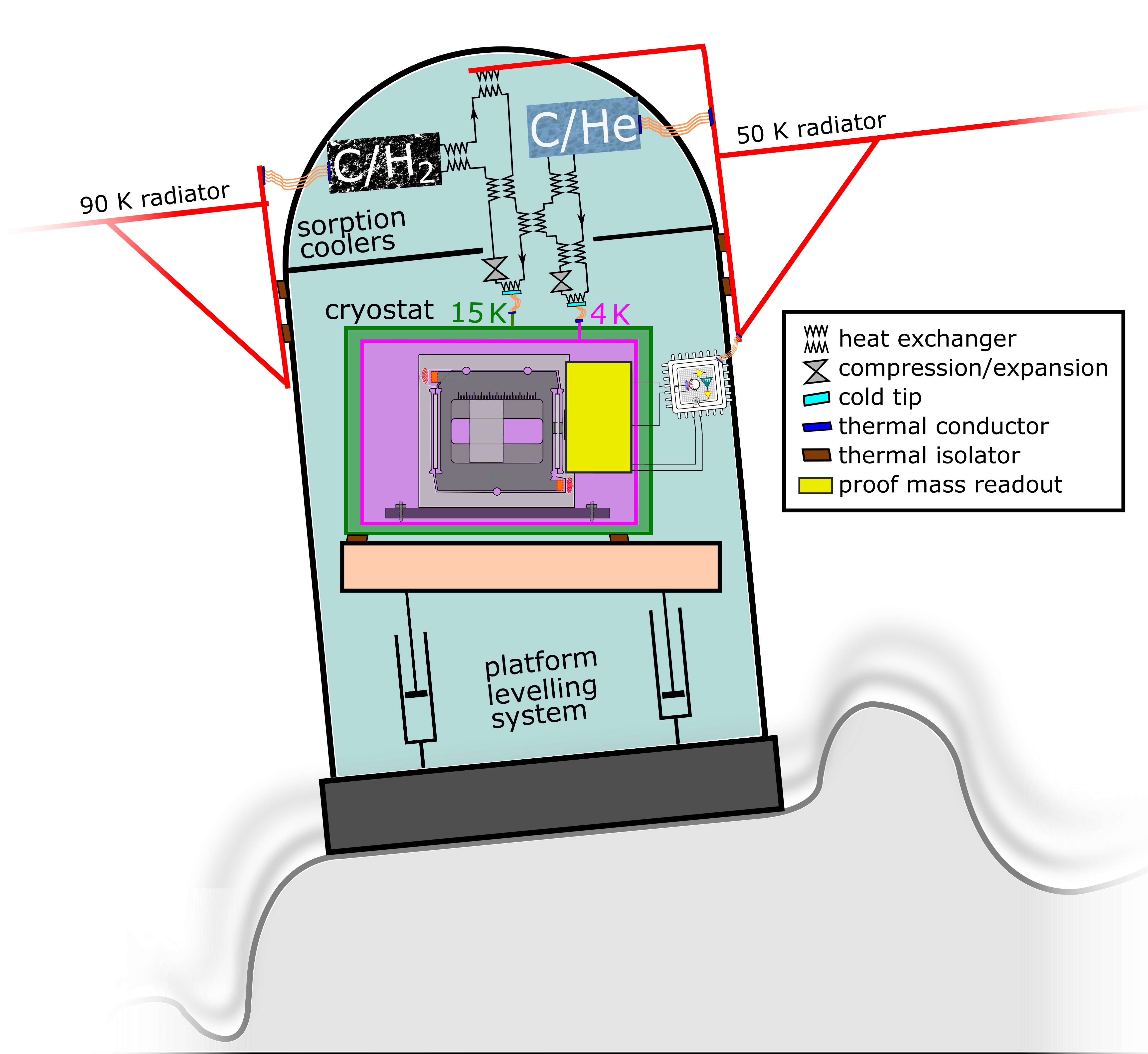}
    \caption{Conceptual overview of a seismic station on a tilted surface on the lunar regolith. Roughness and tilt of lunar surface exaggerated for illustrative purposes. Several subsystems vital to successful operation are depicted and further detailed in the text. Subsystems are not shown to scale here.}
    \label{fig:SeisStation}
\end{figure}

In order to avoid performance limitations from thermal effects, it was proposed to deploy LGWA inside a permanently shadowed region (PSR). The PSRs are formed by craters at the lunar poles. They can have temperatures continuously below 40\,K and be thermally stable with temperature fluctuations driven by heat flow from the lunar interior, infrared light emitted by sunlit parts of the lunar surface, and by scattered sunlight \cite{Glaser2021}. The cold temperatures of a PSR will have the additional benefit to act as a natural cryo-cooler of the proof mass, which lowers thermal noise and enables a sorption-based technology to cool the LGWA proof masses to 4\,K (see section \ref{SorpTherm}). A concept drawing of an LGWA seismic station containing the inertial sensor, a sorption cooler and levelling systems is shown in figure \ref{fig:SeisStation}.

Since it is important to have reliable models of the seismic background for the planning of LGWA, it was proposed to deploy a geophysical explorer mission inside a PSR called LGWA Soundcheck\,\cite{Harms2021b}. The sensitivity target is less ambitious (picometer resolution in the decihertz band), but nevertheless, it will mark a major step forward in lunar seismometer technology and beat the sensitivity of Apollo seismometers by 2 -- 3 orders of magnitude below 1\,Hz. LGWA Soundcheck will allow us to make a greatly improved prediction of the seismic background spectrum based on the observed distribution of seismic events inside a PSR.

\section{Inertial sensor development}\label{Inert}
An LGWA inertial sensor has stringent requirements such as fm/$\surd$Hz sensitivity down to 1\,Hz, deployablility, low heat dissipation and favourable electronic characteristics. While still under development, we describe the current R\&D efforts here. The proof mass will be suspended by means of a folded Watt's linkage, a common way\cite{Bertolini2006} to achieve a compact, low-resonance-frequency device. To achieve low thermal noise, the target proof mass will be 10\,kg. By using niobium, which has a 8.4\,g/cm$^3$ density, such device with all auxiliary sensing and actuation system can fit in a volume 200$\times$200$\times$100\,mm$^3$.  

The readout of the proof mass motion, and therefore ultimately the differential signal between the elastic response of the Moon to passing GWs and the inertial proof mass which holds the GW signal, is a cm-scale interferometer. An example of such opto-mechanical device is a room temperature version of an interferometrically Watt's linkage that reached 8\,fm/$\surd$Hz from 30\,Hz\,\cite{Heijningen2018}. The used interferometric readout, based on ref.\,\cite{Gray1999}, reached 4\,fm/$\surd$Hz from 4\,Hz onwards\,\cite{Heijningen_PhD}. This readout needs feedback to keep the working point halfway up the fringe (the linear part of the sinusoid) as any deviation makes the output non-linear and degrades the subtraction of common mode noise between the two interferometer output ports. Without feedback the typical micrometer motion on Earth of the sensor frame would cause the sinusoidal error signal to move between fringes.

The feedback is provided by an actuator that \textit{locks} the proof mass to the suspension frame. The signal sent to the actuator is then proportional to force and acceleration and serves as the sensor output. Often, a coil-magnet actuator is used in force-feedback inertial sensors. However, in the previously discussed 8\,fm/$\surd$Hz results, thermal noise was expected to be dominant below 10\,Hz. While the used Watt's linkages can have mechanical quality factors above 5000, the permanent magnet and its eddy current damping of the moving metal pieces had degraded the $Q$ to below 100\,\cite{Heijningen_PhD}. LGWA requires lower-frequency fm/$\surd$Hz sensitivity which can only be obtained by lowering thermal noise which goes as\,\cite{Saulson1990}
\begin{equation}
x_{\mathrm{th}}^2 = \frac{4 k_{\mathrm{B}}T\omega_0^2 \phi}{m\omega \left[(\omega_0^2-\omega^2)^2 + \omega_0^4 \phi^2 \right] },
\end{equation}
where $x_{\mathrm{th}}$ denotes the thermal noise displacement amplitude spectral density (ASD), $k_{\mathrm{B}}$ Boltzmann's constant, $T$ the temperature, $\omega_0$ the angular resonance frequency, $\phi (= 1/Q$ for structurally damped suspensions) the loss angle and $\omega$ the angular frequency. Low temperatures and increased mass will obviously help, but different actuators that will not (dominantly) damp the Watt's linkage are necessary. Therefore, superconducting actuators that use the Meissner effect rather than a magnet to exert a force on the proof mass are investigated\,\cite{ElvisCoils,Heijningen2022}. The superconducting thin film coils and superconducting surface (depicted by orange rectangles) can, depending on the achieved cooling level or other application, be manufactured from niobium ($T_{\rm{c}}$ = 9.2\,K), MgB2 ($T_{\rm{c}}$ = 40\,K) or
YBCO ($T_{\rm{c}}$ = 93\,K). To be in the necessary full magnetic expulsion state, temperatures around 60\% of Tc or lower is needed. 

\begin{figure}[h]
  \centering
    \includegraphics[width = 1\linewidth]{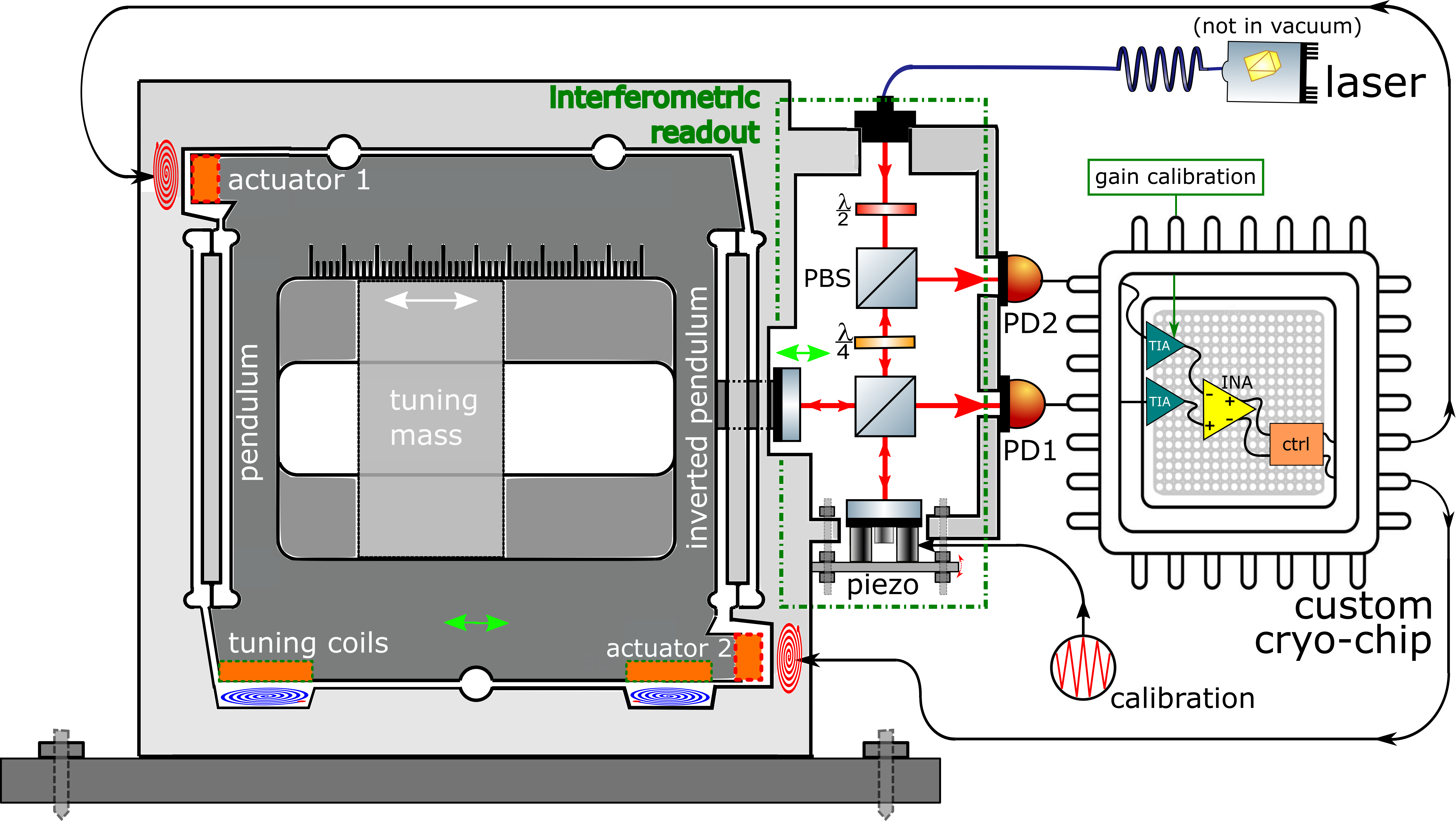}
    \caption{A cryogenic superconducting monolithic inertial sensor. The proof mass is suspended from the frame by a regular pendulum and inverted pendulum. This monolithic configuration is known as a Watt’s linkage and allows for an arbitrarily low natural frequency, which increases the mechanical sensitivity. The proof mass motion is monitored by an interferometric readout and the custom cryo-chip is under development using 65\,nm CMOS technology. More details are found in the text.}
    \label{fig:CSISoverview}
\end{figure}

The current design follows from an initial
cryogenic inertial sensor concept  first proposed in ref.\,\cite{Heijningen2020}, which was subsequently updated\,\cite{Heijningen2022}. Currently, we investigate what is depicted in figure\,\ref{fig:CSISoverview}. The resonance frequency of the Watt's linkage can be coarsely set by the sliding tuning mass, which changes mass distribution between inverted and regular pendulum. After cooldown of the mechanics, the resonance frequency may have changed. A DC current on one of the tuning coils can effectively change the mass distribution thereby tuning the resonance frequency.

The estimate of sensitivity is made by modelling the displacement noises of mechanical and interferometric nature. Most models for these noises are described in refs.\,\cite{Heijningen2018,Heijningen2020}. The actuator noise model is a simple current driver model\,\cite{Heijningen_PhD}. We use the parameters in table\,\ref{table:parameters} and arrive at the noise budget shown in figure\,\ref{sfig:NbIFO}. This noise budget is roughly the same as the "opto-mechanical" trace in figures 2, 3, 4 and 5 of ref.\,\cite{LGWA}. The sensitivity of the four-sensor array is a factor 2 lower. LGWA sensitivity is obtained by dividing out the Moon's response, i.e. the expected surface motion per unit strain. An example of such modeled response is found in figure 1 of ref.\,\cite{LGWA}.

\begin{table} [h]
 \caption{Mechanical, readout and electronics parameters for both the interferometrically and SQUID read out Watt's linkage.}\label{table:parameters}
  \centering
 \begin{tabular}{  l  c  c  }
Parameter & Value & ~~~Unit~~~ \\
   \hline
   \hline

Proof mass & 10 & kg \\
Natural frequency & 0.25 & Hz \\
Temperature & 5 & K \\
Coil-superconductor gap & 0.1 & mm \\
Actuator strength & 50 & \textmu N/A \\
~ & ~ & ~ \\
Niobium with interferometric readout & ~ & ~ \\
\hline
Watt's linkage material & Nb & - \\
Quality factor & 1$\cdot 10^{4}$ & - \\
Frequency noise\footnote{Typical value for high-end lasers e.g. The Rock$^{\rm{TM}}$ from NP Photonics\,\cite{RockSpecs}.} & 500 $\cdot~f^{-1/2}$ & Hz/$\surd$Hz \\
Static differential arm length & 0.5 & mm \\  
Injected laser power & 10 & mW \\
Wavelength & 1550 & nm \\
TIA feedback resistor & 20 & k$\Omega$ \\
 ~ & ~ & ~ \\

Silicon with SQUID readout & ~ & ~ \\
 \hline
Watt's linkage material & Si & - \\
Quality factor & 1$\cdot 10^{6}$ & - \\
SQUID energy resolution $E_{\rm{A}}$& 2500 $\hbar$ & J/Hz \\
signal to SQUID coupling efficiency $\eta\beta$ & 0.25 & - \\
1/$\surd$f corner frequency $f_{\rm{c}}$& 0.1 & Hz \\ 
  \hline

  \hline 
\end{tabular}
 
\end{table}

The used readout scheme is an example femtometer-class interferometer. There are other options to realise an optical readout with similar or even lower predicted sensitivity. The trade-off between displacement readout schemes relies heavily on the required dynamic range and the ability, and corresponding benefits, of operating at a specific or a random operating point. So-called multi-fringe interferometric sensors implement 
\textit{phasemeters} to read out the phase at any operating point and with large, mostly multi-fringe, dynamic range \cite{watchi2018}. These types of interferometers are limited to femtometer-level sensitivities by effective technical-fundamental limitations in their readout, especially by digitisation noise and to provide linear sensing over a wide range they typically do not employ optical resonators to enhance the signals \cite{eckhardt2022}. 

The best space-based demonstration of such displacement sensors is the multi-fringe heterodyne interferometry realised in LISA pathfinder \cite{armano2021a}, which achieved a displacement measurement noise floor  of 30\,fm/$\surd$Hz around 1\,Hz, mostly limited by ADC quantisation noise in the digital phasemeter. A lower digitisation noise floor could be realised with commercially available ADCs. A critical part of the low-frequency noise floor that has to be evaluated for LGWA is the achievable temperature stability and the corresponding thermally driven couplings, namely thermoelastic and thermorefractive noise, which were suppressed in LISA Pathfinder by the exceptional temperature stability \cite{Armano2019}. These thermally driven noise sources will be critical for any interferometric readout scheme and need to be studied with respect to the cryogenic environment of the proof mass. Thermal compensation strategies can be employed, but are complicated, in design  and in testing, by the cryogenic operating temperatures. These noise source are also critical for any opto-mechanical laser frequency reference, be it a proper 2nd, equally long, arm in the local interferometer topology or some external, disjoint reference.

For the LGWA and especially LGWA Soundcheck the power consumption of the payload might be a critical factor, with the laser sources being a significant driver of such a budget. Accordingly, the power consumption of any given interferometric readout has to be taken into account, as well as their influence on the potentially reduced power consumption in the active feedback to control the proof mass. This might benefit interferometric readout schemes that require little or no opto-electronic elements, slow signal digitisation and little signal post-processing. In addition to the readout scheme shown in figure~\ref{fig:CSISoverview}, a higher dynamic range option that can achieve femtometer-level displacement noise with no additional active components is quadrature homodyne interferometry, which has already been used to demonstrate compact interferometric readout of inertial sensors\,\cite{cooper2021} and demonstrated a noise floor of 20\,fm/$\surd$Hz\,\cite{cooper2018}. Depending on the dynamic range and the optical design, especially with regards to ghost beams and polarisation contamination \cite{gerberding2021}, such a readout might require additional digital signal processing with a Lissajous fit to suppress periodic non-linearity, which again might limits its advantage in terms of power consumption.

\begin{figure}[h]
  \centering
    \includegraphics[width = 1\linewidth]{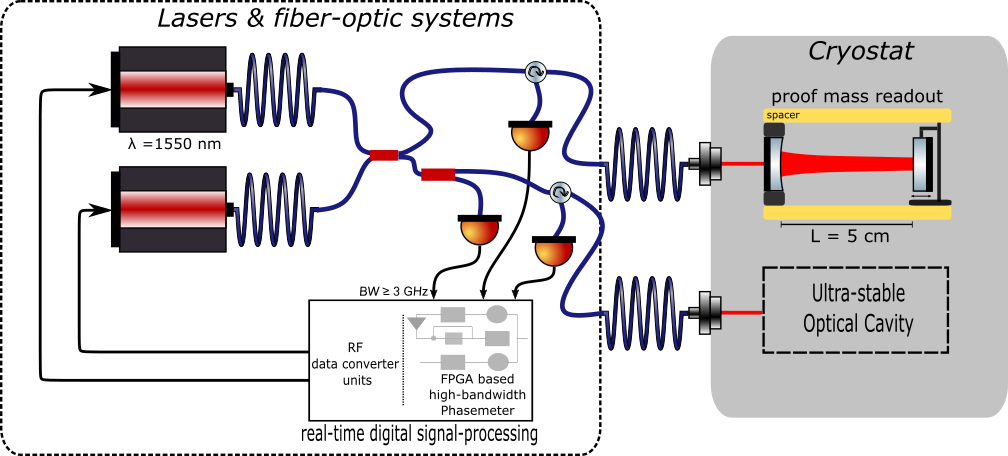}
    \caption{Heterodyne cavity-tracking readout scheme with co-located ultra-stable optical cavity. Tracking the motion of the proof mass requires a high-dynamic range phase readout system. Cavity length $L$, wavelength $\lambda$ and phase readout bandwidth $\text{BW}$ determine the maximum one-way displacement tracking range $\Delta L _{\mathrm{max}} = \lambda/2 \cdot \text{BW} / ( c / ( 2 L) )   = \lambda/2 \cdot \text{BW}/\text{FSR}$. }
    \label{fig:HetCavReadout}
\end{figure}

Finally, optical resonators can be employed in compact displacement sensors to achieve sub-femtometer displacement readout noise floors at the cost of readout range and linearity, for example using fiber-based implementations, as demonstrated in ref.\,\cite{cervantes2014}. Combining optical cavities with operation-point independent, wider-range readout is, however, non-trivial. Using a strong frequency-modulated laser with an optical resonator promises noise floors of $10^{-16}$\,m/$\surd$Hz  \cite{eckhardt2022}, but might require too much effort with respect to opto-electronics and signal processing for the readout of only two displacements in a single LGWA station. A more relevant approach might be to lock one laser to an optical cavity between the proof mass and an external mirror and to measure its frequency variations with changing length. Such a scheme requires a second ultra-stable laser to generate a beat note, but, combined with a corresponding real-time digital signal-processing system, this scheme can also realise the locking of both lasers to their respective optical resonators \cite{eichholz2015}, as depicted in figure\,\ref{fig:HetCavReadout}. 

This readout senses one degree-of-freedom, adding another axis demands an additional laser that is locked to the corresponding cavity. Hence, in order to measure the horizontal surface displacement along two orthogonal directions, each seismic station requires in total three laser sources.
For resonator lengths of 5\,cm the beat frequency will shift by 3\,GHz for a displacement of $\lambda/2$, a frequency shift that could be tracked with a high-bandwidth, frequency-tracking phasemeter\,\cite{Gerberding2013} with negligible frequency tracking noise. Field-programmable gate arrays with integrated high-speed data converters are available to implement such tracking systems with several GHz of bandwidth. A heterodyne cavity-tracking readout scheme can, in principle, achieve readout noise levels of $10^{-17}$\,m/$\surd$Hz with reasonable levels of cavity Finesse, because they are not directly limited by digitisation noise and the influence of shot-noise is suppressed by the optical enhancement. In practise this readout will be limited by the stability of the available frequency reference, which could be a separate cavity as developed for space-based optical clocks or fundamental physics experiments \cite{gurlebeck2018} that is co-located within the cryostat to reduce thermal effects like coating thermal noise, as shown in figure\,\ref{fig:HetCavReadout}. If available, the ultra-stable laser can also be a fully separate device connected only via fiber. The lasers, the phase readout system and the fiber-optics do, to first order, not have stringent environmental noise couplings and can be placed outside the cryostat.
The additional complexities and power consumption of a heterodyne cavity locking scheme make it unsuited for LGWA Soundcheck, but the promise of mid-range dynamic range and extremely low readout noise floor make it a promising candidate for the full LGWA readout, assuming other noise sources can be brought to sufficiently low levels, at least at the higher readout frequencies. Detailed studies of amplitude noise \cite{wissel2022} and of tilt-to-length coupling \cite{hartig2022} will have to be done for any design and readout scheme.

\begin{figure}[h]
\centering
	\subfigure[~]{
	\includegraphics[width=0.47\textwidth]{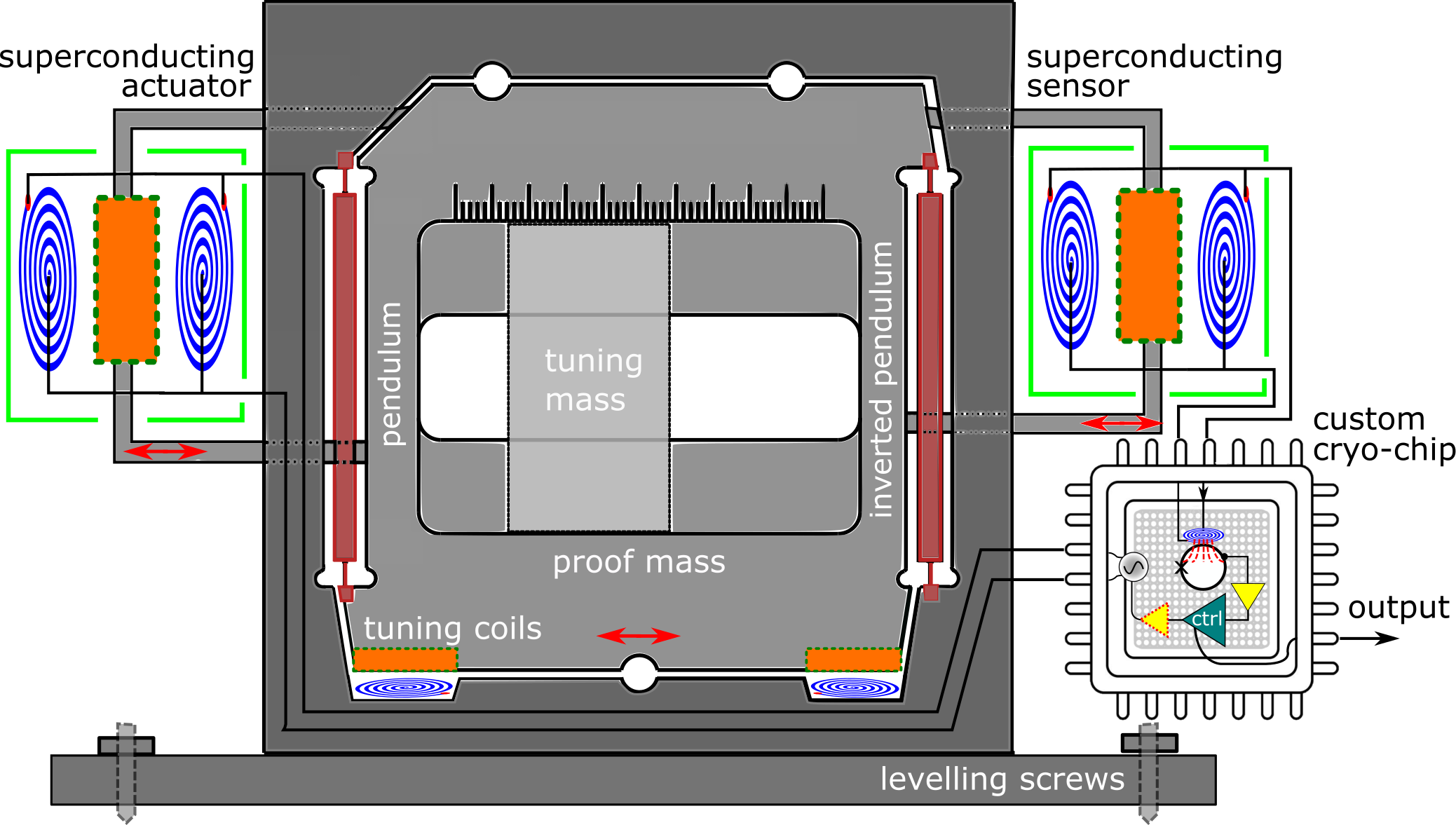}
    \label{sfig:LGWA_inertial_SQUID}}
	\subfigure[~]{
	\includegraphics[width=0.35\textwidth]{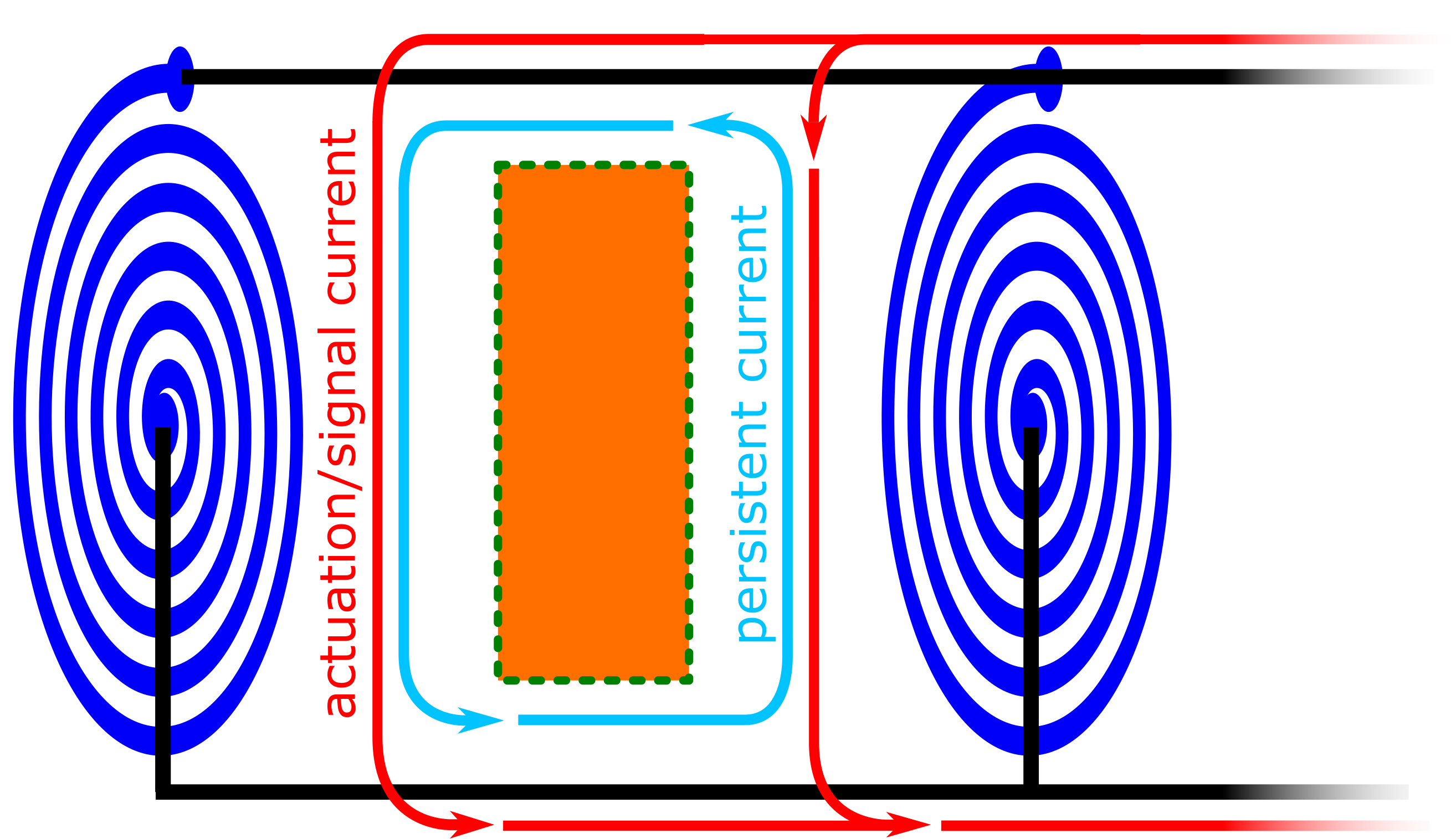}
    \label{sfig:TwoCoilSandwich}}	\subfigure[~]{
	\includegraphics[width=0.43\textwidth]{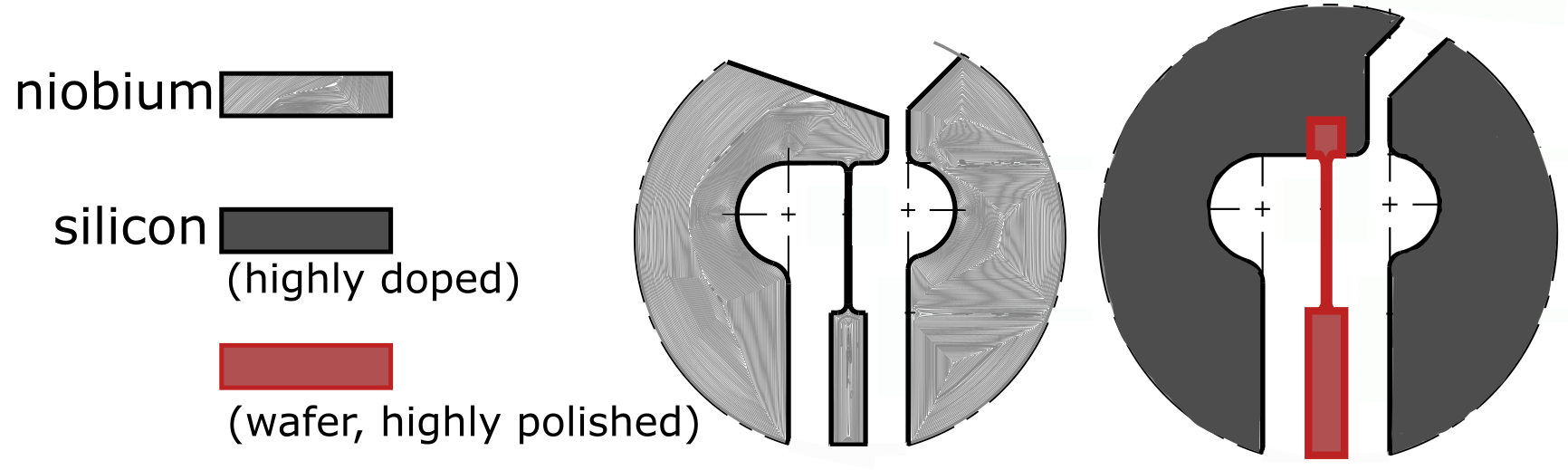}
    \label{sfig:NbSiFlex}}
\caption{  \subref{sfig:LGWA_inertial_SQUID} a silicon Watt's linkage with superconductive readout and actuation, \subref{sfig:TwoCoilSandwich} dual coil sandwich configuration used for sensing and actuation. More details found in text and \subref{sfig:NbSiFlex} a zoom of the monolithic niobium and quasi-monolithic silicon flexures} 
\label{fig:futureCSIS}
\end{figure}

\begin{figure}[h]
\centering
	\subfigure[~]{
	\includegraphics[width=0.47\textwidth]{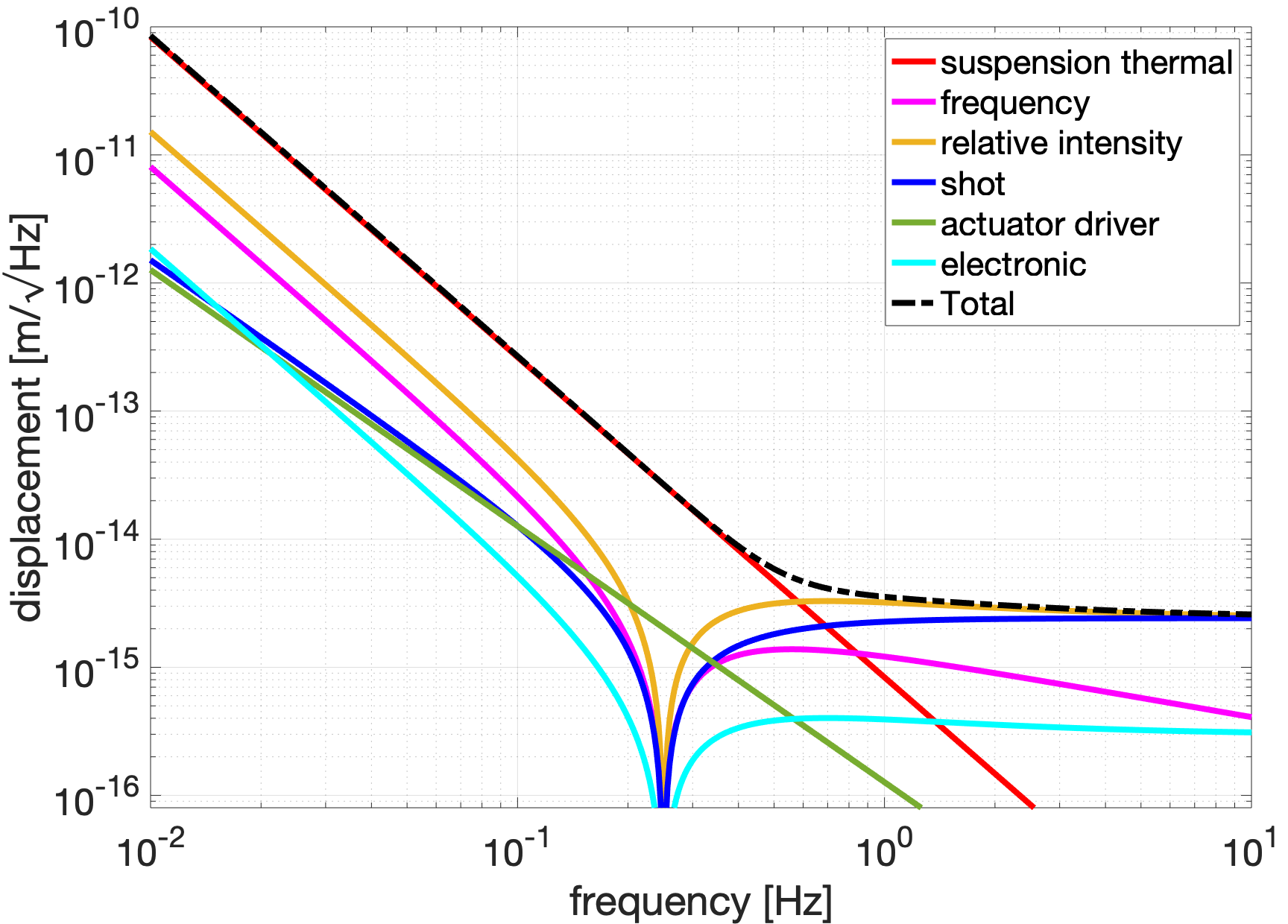}
    \label{sfig:NbIFO}}
	\subfigure[~]{
	\includegraphics[width=0.47\textwidth]{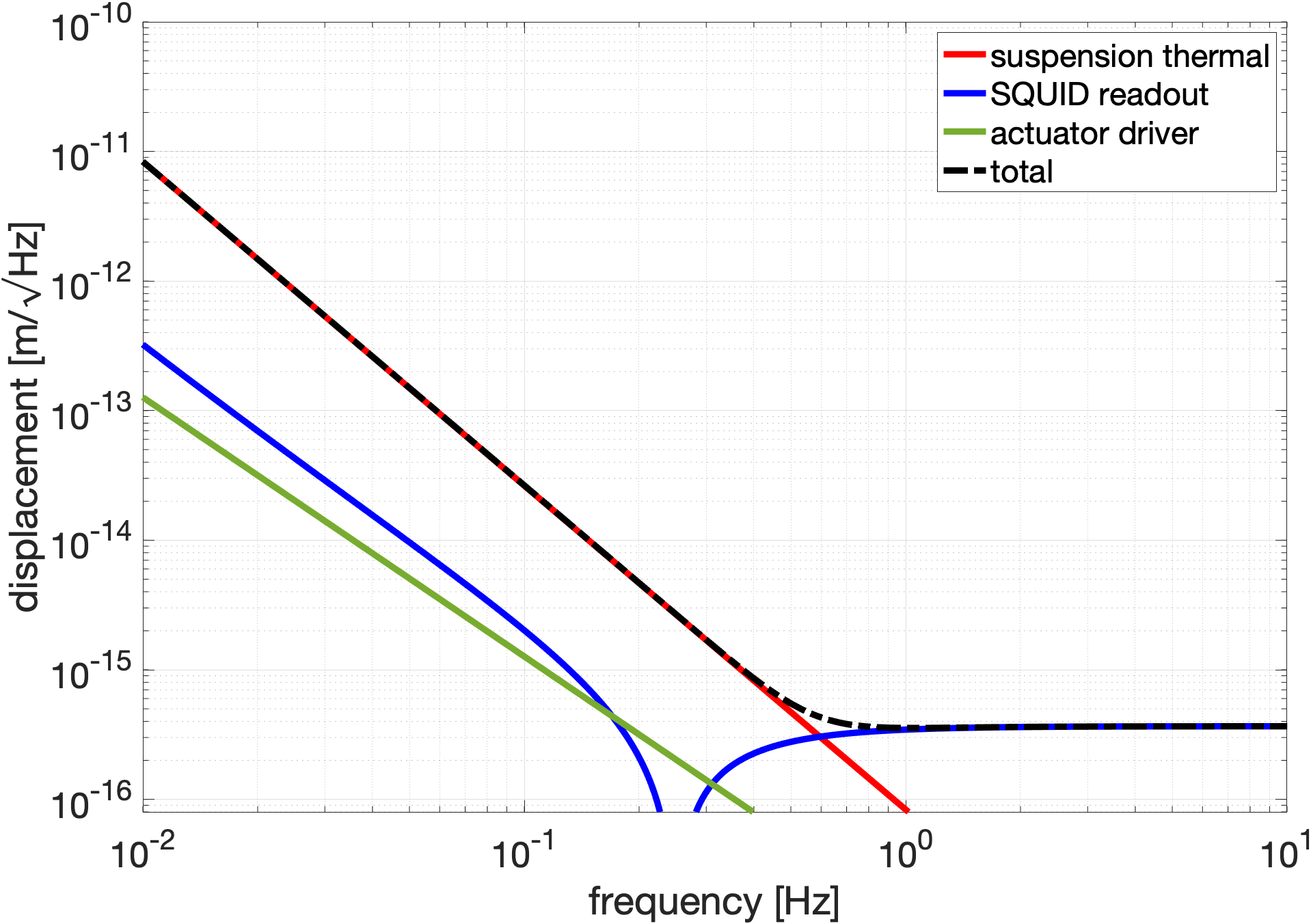}
    \label{sfig:SiSQUID}}
\caption{Minimum detectable inertial  displacement for a structurally damped accelerometer with \subref{sfig:NbIFO} niobium  mechanics and interferometric readout and \subref{sfig:SiSQUID} silicon mechanics and SQUID readout.}
\label{fig:LGWAsens}
\end{figure}

Besides the different interferometric readout strategies described above, superconductivity can be used to read out the proof mass position with high precision. If a superconductor moves with respect to a superconducting coil carrying a persistent current, the inductance of coil-superconductor system changes. The current in the coil will change correspondingly to keep the ﬂux in the system conserved due to ﬂux conservation in superconducting loops. The current change can beconverted to magnetic ﬁeld change simply by connecting another coil. This changing magnetic field can subsequently be picked up by a Superconducting QUantum Interference Device (SQUID), which is known for its extreme sensitivity to changing magnetic fields. This readout strategy has been suggested, e.g., in ref.\,\cite{Paik1976} for gravity gradiometry. Using two sensing coils in parallel and sandwiching a superconductor, and a third one to convert the current signal into magnetic signal, the motion of the superconductor can be read out with sub-femtometer precision. On the right side of figure\,\ref{sfig:LGWA_inertial_SQUID} such dual coil sandwich configuration is shown. 

The superconducting readout provides an error signal for a feedback loop with a superconducting actuator, which can also employ a dual coil sandwich architecture. The superconducting coils can be loaded with a persistent current as shown in figure\,\ref{sfig:TwoCoilSandwich}. By sending an actuation current running in parallel in the two coils, we can increase the current, and corresponding magnetic force, on one side and reduce the magnetic force on the other side, generating a net (feedback) force on the superconductor. The magnetic force between a coil and a superconducting surface is proportional to the square of the current in the coil $(I_{\rm{pers}}+I_{\rm{act}})^2 =I_{\rm{pers}}^2+2I_{\rm{pers}}I_{\rm{act}}+I_{\rm{act}}^2$. Large persistent currents (> 1 A currents are common\,\cite{CollPaikPC2022}) will give the largest
coupling to the signal current. However, because the persistent currents in the coils push from either side there is a positive stiffness roughly equal to the DC force from each coil, divided by the coil-surface gap. This added stiffness can be corrected for using the tuning mass and coils. The main advantage of this strategy is that only small currents (< 100 \textmu A) will have to be generated by the on-chip current driver. Moreover, the dual coil architecture linearizes the relation between the actuation current and the feedback force which will simplify the control and data analysis. 

To decrease the thermal noise even further, a silicon Watt's linkage is proposed. Silicon is a crystalline material exhibiting low mechanical loss at cryogenic temperatures, with a bulk $Q$ of $10^8$\,\cite{Nawrodt2008}. The thin flexures allowing for their low stiffness of metallic Watt's linkages have historically been fabricated using electro-discharge machining (EDM) techniques as shown in figure\,\ref{sfig:NbSiFlex}. A more difficult hybrid
procedure for silicon must be followed as using EDM to cut the delicate flexures is expected to result in surface damage and thus lossy flexures. The frame and proof mass are manufactured from highly doped silicon,
which can be cut using EDM. The legs including the flexures are (laser assisted plasma) etched out of a thick
500 \textmu m wafer and hydro
catalysis bonded (HCB) to the frame and proof mass. HCB is famous for producing quasi-monolithic bonds in mirror suspensions of the current interferometric GW detectors\,\cite{vanVeggel2014}. Figure 2 in ref.\,\cite{ElvisCoils} shows a possible HCB assembly procedure for a silicon Watt's linkage. The quasi-monolithic silicon Watt's linkage is expected to have a $Q$ of $10^6$, thereby lowering the thermal noise by an order of magnitude with respect to the niobium variant.

The SQUID readout has a sub-fm/$\surd$Hz sensitivity corrected for the sensor mechanics as\,\cite{PaikVeth2009}
\begin{equation}
    x_{\rm{squid}}^2 = \frac{2 E_{\rm{A}}(1+f_{\rm{c}}/f)}{m \omega_0 \eta\beta} \frac{(\omega^2-\omega_0^2)^2+\omega_0^2/Q}{\omega^4}, 
\end{equation}
where most symbols have been denoted in table\,\ref{table:parameters}. The SQUID has a 1/f characteristic below $f_{\rm{c}}$ in its power spectral density. The same actuator noise model as the niobium version and the silicon proof mass suspension thermal noise model complete the noise budget as presented in figure\,\ref{sfig:SiSQUID}.

\section{Sorption cooling and thermal management}\label{SorpTherm}

Cryogenic cooling of the inertial sensor will be established by combining two vibration-free cooling technologies; High-emissivity radiator panels will be used to provide heat-sink platforms at temperature levels of about 50\,K and 90\,K. Next, a two-stage sorption-based Joule-Thomson cooler will be heat sunk to these platforms and will cool further down to 14.5\,K and 4.5\,K. This sorption-based cooling technology has been developed at the University of Twente in the past two decades. It operates with a thermal compressor rather than a mechanical compressor as conventional cryogenic coolers do.  Apart from a few passive valves it has no mechanical moving parts and, therefore, offers operation at an extremely low level of emitted vibrations and a long lifetime because of the absence of wear. Both aspects are obviously attractive in space applications. The operation of a sorption compressor is based on the cyclic adsorption and desorption of a working gas at a sorber material such as, in our case, activated carbon. Activated carbon is a material that by its highly porous structure has a very large internal surface so that it can adsorb large quantities of gas. By heating the sorber, the gas is desorbed and a high pressure can be established. By expanding this high-pressure gas in a Joule-Thomson (JT) cold stage, cooling can be obtained.  The operating principles and the thermodynamics involved, are discussed in many papers\,\citeg{Burger2002,Burger2007,Doornink2008,Wiegerinck2006,Wiegerinck2007}.

\begin{figure}[h]
\centering
	\includegraphics[width=0.4\textwidth]{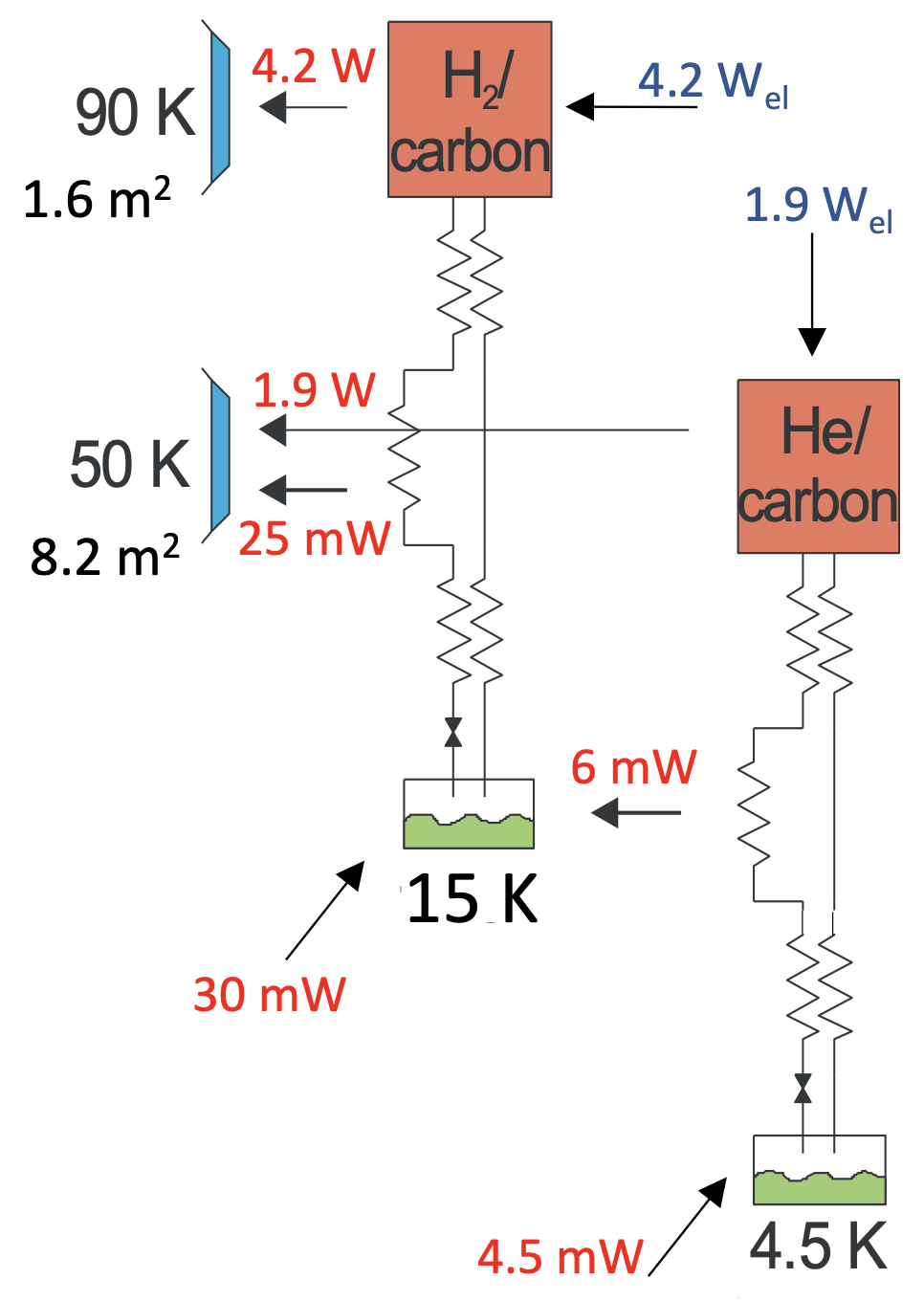}
\caption{Schematic diagram of the two-stage sorption-based cooler with a cooling power of 30\,mW at 15\,K and 4.5\,W at 4.5\,K; Electric input power is indicated in blue; heat flows in red.}
\label{fig:SorpWatt}
\end{figure}

The baseline cooler chain of the LGWA project is schematically depicted in figure\,\ref{fig:SorpWatt} and resembles the Darwin cooler that was developed in an earlier ESA-TRP project\,\cite{Burger2007}. The first stage of the LGWA sorption cooler operates with hydrogen gas and realizes a temperature of 15\,K. The second-stage sorption cooler operates with helium gas and, precooled by the hydrogen stage, it reaches 4.5\,K. The hydrogen compressor is thermally linked to the 90\,K radiator heat sink. The hydrogen gas is precooled by a 50\,K radiator that also serves as the heat sink for the helium compressor. Based on the performance of the two stages of the Darwin cooler, the gross cooling powers at both stages in the LGWA project are expected to be 36\,mW at the 15K stage (of which 6\,mW are used to precool the helium gas in the second stage), and 4.5\,mW at the 4.5\,K stage. The total electric input power to the coolers is slightly more than 6\,W; 4.2\,W in the compressor of the hydrogen stage and 1.9\,W in that of the helium stage. This input power, plus the power taken from the cold interfaces is emitted to deep space at the two radiator panels. In previous work, the radiator temperatures were optimized aiming at minimum radiator size, resulting in actual temperatures of 87\,K and 51\,K. The required radiator panel areas are 1.6\,m$^2$ and 8.2\,m$^2$, respectively\,\cite{terBrake2011}. This setup is schematically depicted in figure\,\ref{fig:SorpWatt}. The cooler mass is expected to be 10 kg of which both stages are about half of that\,\cite{Burger2007,terBrake2011,ESAcontract2007,ESAcontract2013}. The cooling powers as indicated in figure\,\ref{fig:SorpWatt} are not fully available as net cooling power. Part of it is used to take up parasitic heat loads due to conduction and radiation. The heat load budgets are listed in table\,\ref{table:heatbudget}. In order to withstand launch loads, all frames will be mechanically fixed. Once positioned on the moon surface, these launch-load connections will be disconnected allowing for the 15\,K frame to be leveled with respect to the moon surface, as illustrated in figure\,\ref{fig:SeisStation}. The remaining support structures are anticipated to be G10 struts between leveling platform and 15\,K frame, and Kevlar straps between 15\,K frame and 4.5\,K cold mass.

\begin{table}[h]
\begin{center}
\label{table:heatbudget}

\begin{tabular}{ l c }
\\
15\,K & ~~ \\
 \hline \hline
Total gross cooling power &	36\,mW \\
  \hline
Precooling He stage & 6\,mW \\
Radiation from 50 K environment	& 20\,mW \\
Conductive load through support (G10 struts)	& 9\,mW \\
Conductive load via cooler tubing & 1\,mW \\
Emissivity & 0.1 \\
\hline
~ & ~ \\	
4.5\,K & ~ \\	
\hline \hline
Total gross cooling power & 4.5\,mW \\
\hline
Radiation from 15 K environment	& 0.1\,mW \\
Conductive load through support (Kevlar straps)	& 0.9\,mW \\
Conductive load via cooler tubing & 0.1\,mW \\
Dissipation and conductive load of sensor + electronics & 3.4\,mW \\
Emissivity & 0.1 \\
\hline
\end{tabular}
\caption{Heat load budgets at the 15\,K and 4.5\,K cold-tip interfaces.}
\end{center}
\end{table}

A sorption-based Joule-Thomson cooler has been launched and successfully operated in space in the ESA-Planck mission (2009-2013)\,\cite{Morgante2009,Planck2011}. It provided cooling power of 1\,W at about 20\,K using hydrogen as the working fluid. However, the compressor sorber material was a metal hydride which is a chemical absorber whereas in our compressor technology activated carbon is applied which is a physical adsorber. The big difference is that a chemical absorber degenerates over time limiting the lifetime of the cooler in mission (in Planck 2 years), whereas the adsorption process with carbon is fully reversible and does not limit the lifetime of the cooler. Our carbon based sorption compressor technology was qualified at TRL5 (surviving launch vibrations) in one of the recent ESA projects\,\cite{ESAcontract2017}.

\section{Seismometer leveling system}\label{Level}
A leveling system is needed to achieve an initial alignment of the seismometer platform to compensate ground slope and then to keep it aligned within a few microradians. The requirement of the alignment accuracy is set by the softness of the proof-mass suspension through the tilt-to-horizontal coupling $d_{\rm{pm}} = g \theta / \omega_0^2$. The critical dimension in figure~\ref{fig:CSISoverview} and~\ref{sfig:LGWA_inertial_SQUID} is the 100\,\textmu m gap between coils and superconductor in the actuator. The leveling system should be more precise than 30\,\textmu m in proof-mass positioning to ensure that the superconductor does not make contact with the sandwiched coils. 

A platform meeting similar requirements was developed for the SEIS experiment of the Mars InSight mission \cite{FlEA2018,BiEA2019,Lognonne2019}. This system features a MEMS-based rough alignment to compensate for up to 15$^\circ$ of ground slope, and a precision alignment system that reaches a few microradians using high-precision tiltmeters. An important new requirement for the LGWA platform is that it must be compatible with the cold environment of a PSR, which constraints above all the technologies that can be used for the high-precision tiltmeters. 

An alternative to using high-precision tiltmeters might be to realize the LGWA seismic sensors with a high dynamic range laser-interferometric readout of the proof-mass displacement \cite{EcGe2022}. Exploiting the tilt-to-horizontal coupling, tilt can be measured and compensated by observing the movement of the proof mass. With the rough tilt alignment stage, one can assess what sign the high-precision adjustment must have, i.e., with which side of its frame the proof mass makes contact before the fine-alignment is engaged.

\section{Synergy with Einstein Telescope}\label{Syn}
On Earth, ET features an underground and cryogenic design and aims to be sensitive to GWs down to 3\,Hz. Methods to apply low-vibration cryogenic cooling of the mirrors in a cryostat to lower thermal noise are currently investigated in research facilities\,\cite{DiPace2021,Utina2022,ETESTinprep}. Close to the mirror spurious vibrations could be injected by the application of cooling power. To ensure the lower cryogenic stages are indeed at low enough vibration levels, new inertial sensors such as described here are necessary. 

\begin{figure}[h]
\centering
	\includegraphics[width=0.55\textwidth]{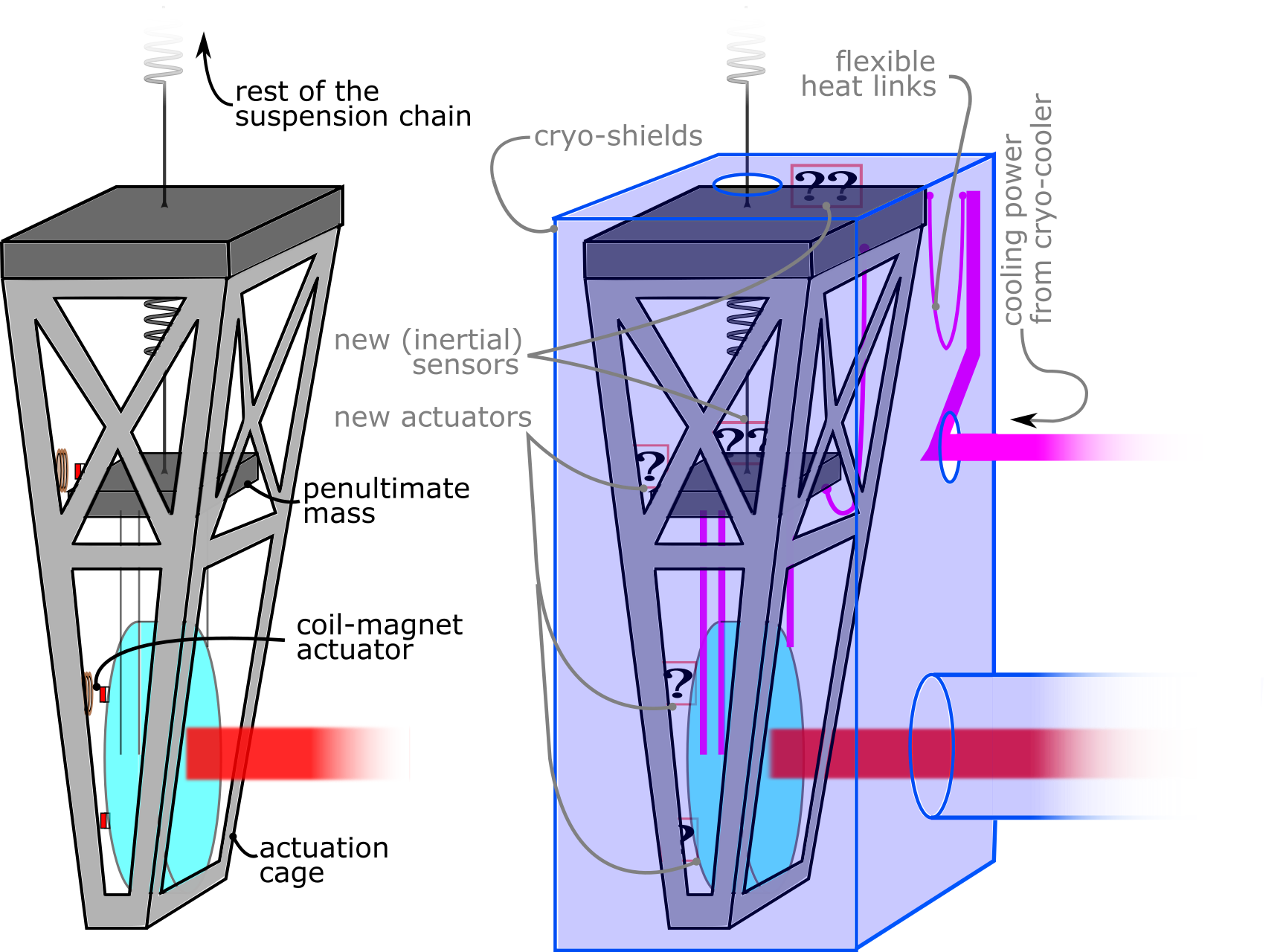}
\caption{The final stages of (left) current room-temperature mirror suspensions and (right) future cryogenic mirror suspensions, where the low temperatures provide opportunities for new actuators and (inertial) sensors. Ultimate configuration for ET may differ; however, similar sensing and actuation solutions will be necessary.}
\label{fig:SuspsRTandCRYO}
\end{figure}

ET aims to be 10 times more sensitive than current detectors above 10\,Hz and stretch its lower bandwidth limit down to 3\,Hz. Cooling down of the input and end mirrors down to around 10\,K is needed to reduce the dominant noise at low frequency: thermal noise. To extract heat, the penultimate mass above the mirror shown in figure\,\ref{fig:SuspsRTandCRYO} (right) operates at about 5\,K. Cooling the penultimate mass cannot be done radiatively due to the low temperature and required power (several 100\,mW) and therefore some physical connection between cryocoolers and the suspension final stages is required. The cooling power is applied by low-vibration cryocoolers and using flexible heat links. However, there is still a risk that unwanted vibrations end up in the penultimate stages, close to the mirrors where extremely tiny displacements in the detection bandwidth are required. The cryogenic temperatures provide opportunities for new, superconductive actuators and (inertial) sensors. The use of superconductive coils reduces the cooling power (and therefore vibrations) otherwise needed for dissipative elements, such as the resistive copper actuator coils in figure\,\ref{fig:SuspsRTandCRYO} (left). Extremely sensitive inertial sensors, such as presented here, are needed to monitor the platform motion.

In GW detector suspensions, actuators are used in an hierarchical way in terms of strength and range; the most low-noise, short-range actuators are needed close to the mirror where residual acceleration is extremely small. At the top of the suspension chain, actuation noise requirements are less stringent, but those actuators will have to operate over a larger range. Most actuators used in today’s GW detectors are (some form of) coil-magnet actuator as these are easy develop, install and use. The use of permanent magnets close to moving metals can cause harmful eddy currents and stray magnetic field can exert unwanted forces on the suspended objects. The former is largely solved by using plastics (e.g. PEEK) near the magnets and the latter is often solved by placing the magnets on the same object in opposite polarity.

The cryogenic GW detector KAGRA operates 23\,kg mirrors dissipating 0.5\,mW\,\cite{Ushiba2021} at the actuators and ET mirrors are 10 times as massive\,\cite{ET}, thus dissipating >10 mW if old resistive actuators are used. This is of order 10\% compared to the expected absorption of laser light and thermal radiation of mirror and payload, respectively. Lastly, the sub-fm/$\surd$Hz dual coil position sensor with SQUID readout can be used as differential sensors between cage and (pen)ultimate stage(s).

\section*{Conclusion and future work}
To open up GW science in the decihertz range, there have been space-borne proposals, such as DECIGO\,\cite{DECIGO} and BBO\,\cite{BBO}. While they promise higher sensitivity than LGWA, many technological challenges remain and a longer timeline is expected. Here, we have presented several technologies that make up the payload and detail several different options in the inertial sensor design.

While the niobium Watt's linkage fabrication processes and interferometric readout technology is more mature, the silicon Watt's linkage with SQUID readout may result in roughly one order of magnitude lower thermal and readout noise. Note that a tenfold sensitivity improvement will lead to larger range and thus an expected factor thousand more  GW signals. In both designs we propose actuators with superconducting coils  which are also necessary for the sensing part in the SQUID readout. The development of the inertial sensor as well as the sensing and actuation technology shows strong synergy with future cryogenic GW detector ET.

The inertial sensors with extreme sensitivity have to be tested in extremely quiet and cold environments. Such test facilities in the form of actively isolated platforms inspired by the LIGO HAM table designs\,\cite{Matichard2015} are being developed as part of the E-TEST effort in Belgium \,\cite{ETESTinprep,DiPace2021} and the GEMINI facility in the underground National Laboratories of Gran Sasso\,\cite{GEMINI}. The aimed-for sensitivity at 1 Hz is about 5 orders of magnitude smaller than the Earth's seismic motion at that frequency. Placing two or three identical sensors on the isolated platform allows for subtraction of common mode noise using the Wiener filter\,\cite{Harms2020} or three-channel correlation techniques\,\cite{Sleeman2006} resulting in a sensor self-noise measurement.

The technology necessary for LGWA will either be specific development of existing space technology (levelling system, sorption cooler, thermal management systems etc.) or in parallel with terrestrial GW instrumentation R\&D in inertial sensing and active isolation. Future terrestrial GW detector isolation has to stretch to lower frequencies and needs better low-frequency inertial sensors and active isolation performance for that. For a space application as LGWA, however, there will be extra (space) engineering necessary. Before LGWA will fly, the aforementioned LGWA Soundcheck also requires some technology development. Its strategy is to combine technologies that have already flown in space. For instance, elements of the interferometer topology developed for LISA (Pathfinder) can be adopted for the readout of Soundcheck. R\&D for LISA and other space missions will also have overlap with the technologies presented here. In this context, payload technology development continues towards cryogenic, (sub-)fm/$\surd$Hz inertial sensing on the lunar surface for GW detection and lunar geophysics.

\section*{Acknowledgements}
Oliver Gerberding and Shreevathsa Chalathadka Subrahmanya are funded by the Deutsche Forschungsgemeinschaft (DFG, German Research Foundation) under Germany's Excellence Strategy---EXC 2121 ``Quantum Universe''---390833306. Filip Tavernier and Alberto Gatti are funded by internal KU Leuven funds (iBOF-21-084). Filip Tavernier, Alberto Gatti, Christophe Collette, Joris van Heijningen and this research are partially funded by Interreg V-A Euregio Maas-Rijn under the E-TEST project (EMR113). Morgane Zeoli is funded by the Fonds National de la Recherche Scientifique (FNRS) under projet de recherche STELLAR (T.0022.22).

\bibliographystyle{unsrt}
\bibliography{refs}

\end{document}